\documentclass[prl,twocolumn,nofootinbib,longbibliography]{revtex4}
\usepackage[utf8]{inputenc}
\usepackage{amsmath}
\usepackage{amsfonts}
\usepackage{amssymb}
\usepackage{graphicx}
\usepackage{float}
\usepackage{color}

\begin{document}
\title{Role of $\alpha$ and $\beta$ relaxations in Collapsing Dynamics of a Polymer Chain in Supercooled Glass-forming Liquid}
\author{Mrinmoy Mukherjee}
\author{Jagannath Mondal}
\author{Smarajit Karmakar}
\affiliation{Centre for Interdisciplinary Sciences,
  Tata Institute of Fundamental Research, 
  36/P, Gopanpally Village, Serilingampally Mandal,
RR District, Hyderabad, 500019, India}

\begin{abstract}
Understanding the effect of glassy dynamics on the stability of 
bio-macromolecules and investigating the underlying relaxation processes governing 
degradation processes of these macromolecules are of immense importance 
in the context of bio-preservation. In this work we have studied the stability 
of a model polymer chain in a supercooled glass-forming liquid at different amount 
of supercooling in order to understand how dynamics of supercooled
liquids influence the collapse behavior of the polymer. Our systematic computer simulation studies find that apart
from long time relaxation processes ($\alpha$ relaxation), short
time dynamics of the supercooled liquid, known as $\beta$ relaxation 
plays an important role in controlling the stability of the model 
polymer. This is in agreement with some recent experimental findings. 
These observations are in stark contrast with the common belief that
only long time relaxation processes are the sole player. We find
convincing evidence that suggest that one might need to review the 
the vitrification hypothesis which postulates that $\alpha$ relaxations 
control the dynamics of biomolecules and thus $\alpha$-relaxation
time should be considered for choosing appropriate bio-preservatives. 
We hope that our results will lead to understand the primary factors 
in protein stabilization in the context of bio-preservation.
\end{abstract}
\maketitle

\section{Introduction}
Many organisms can survive in dehydrated state for long period of times by 
accumulating large amount of sugars (sometime $20-50$ $\%$ of the dry weight)
\cite{CarpenterCroweCrowe98,review1}. These carbohydrates (mainly trehalose 
and sucrose) stabilize proteins and membranes in dry state \cite{review1}. 
There are many hypotheses for this protein stabilization and they mainly 
focus on the vitrification of the stabilizing sugar matrix along with the 
biomolecules and water replacement from the neighbourhood of the biomolecules 
by the sugar \cite{science1,review1,KDCRIJPharm99}. In the water 
replacement hypothesis, it is believed that water molecules are replaced 
by sugar which provides appropriate hydrogen bonds to polar residues of 
macromolecules thereby stabilizing them thermodynamically. A slightly 
refined hypothesis is water entrapment hypothesis, in which it is argued 
that interfacial waters provide stabilization of local conformations of 
biomolecules. Some regions on the surface of the biomolecule are more 
hydrophilic than others which leads to preferential binding of water 
molecules at biomolecule-sugar interface.

On the other hand, vitrification hypothesis is purely based on kinematics. 
It is assumed that the carbohydrates form glasses at high concentrations 
or in dry state and thereby slow down the degradation process of biomolecules. 
This hypothesis mainly focuses on how glassy materials relaxes at longer time 
scale. A well-known example of such a phenomena is the preservation of insects
in amber for millions of years, suggesting that vitrification is one of 
the best choices for nature for bio-preservation. A recent hypothesis which 
is a variant of the vitrification hypothesis, suggests that the shorter 
time-scale $\beta$ relaxation rather than slower and longer time $\alpha$ 
relaxation of the glass-forming liquid is actually responsible for the 
degradation of the biomolecule in sugar glasses
 \cite{CiceroneDouglasSoftMatter2012, CiceroneDouglasBioPhyJ2004}.

All of these hypotheses suggest rather different approaches to design 
appropriate sugar glass model to optimally increase the stability of 
biomolecules for preservation. A clear understanding towards this direction 
warrants consideration of all the relevant relaxation processes in glass 
forming liquids,which is summarized below
 
 Relaxation of density fluctuations in supercooled liquids is hierarchical 
and happens in multiple steps as the putative glass transition temperature 
is approached. After a fast initial decay the correlation functions approaches 
a plateau and then at subsequent long-time it decay to zero. The relaxation 
that happens in the plateau like regime is called $\beta$ relaxation and 
the longer time decay from the plateau to zero is called $\alpha$ relaxation
\cite{11BB,arcmp,KDSROPP16,SK2016}. It is well-known that the $\alpha$-relaxation
is very heterogeneous and cooperative in nature with a associated growth of 
a dynamic heterogeneity length scale \cite{KDS,arcmp,KDSROPP16}. On the 
other hand $\beta$-relaxation is believed to be more local process without 
any significant growth of correlation length \cite{JG}, but recent studies have 
suggested that shorter time relaxation processes are probably also 
cooperative in nature with length scale that grow very similarly as the 
long time dynamic heterogeneity length scale \cite{betaPRL,footnote}. 

This indicates that if cooperative motions are required for certain relaxation 
process to happen in a molecules embedded in supercooled liquid, then both short 
and long time relaxation processes will probably play equally important role. For 
example, if $\alpha$-relaxation plays a key role in degradation of protein 
molecules in glassy matrix due to its cooperative nature to induce mobility 
in these biomolecules which are much larger compare to the solvent molecules, 
then shorter time $\beta$-relaxation process will also be able to induce such 
mobility especially at lower temperatures where the time scale related to 
$\beta$-relaxation does not become super exponentially larger compare to 
$\alpha$-relaxation time. Indeed in a recent experiment, it is suggested 
that $\beta$-relaxation plays very important role in the preservation of 
protein in sugar glasses \cite{CiceroneDouglasSoftMatter2012, CiceroneDouglasBioPhyJ2004}.

Although the physical and chemical processes 
that degrade a macromolecule is known, the microscopic mechanisms  of how glassy 
matrix helps to slow down these physical and chemical degradation
process of a biomolecules is not clearly understood. A clear understanding of  
these microscopic mechanisms will reduce the trial-and-error aspect of lengthy and 
tedious long-term stability studies in many fields such as food, pharmaceuticals. The goal of this work is to understand how the dynamics of biomacromolecule might couple 
to the dynamics of supercooled liquids and how rates of different processes 
are modified by the embedding liquid as it is supercooled with decreasing 
temperature. In this regard, we have quantified the collapse dynamics of 
a model polymer chain  \cite{BZZJACS2009} in a well-known glass forming 
liquids\cite{KA} using extensive molecular dynamics simulations. The use of a 
homopolymer as a system of choice avoids the inherent molecular heterogeneity 
of diverse amino-acids in a protein, where isolating individual contributions 
to a glassy-matrix induced change in stability is a difficult task and also 
provides an incentive for exploring the action of glassy matrices on 
hydrophobic interaction, one of the central driving forces for protein folding.  
The model helps us to clearly understand how crowding due to the 
dense packing of embedding glassy liquid molecules particularly influence 
the dynamics of model biomolecule and whether the dynamics of the biomolecule 
can be slaved to the dynamics of the supercooled liquids. We find that indeed 
the dynamics of a polymer chain can be slaved to the dynamics of the
supercooled liquid even when the polymer interacts very weakly with the 
liquid molecules.

The rest of the paper is organized as follows. First we will discuss about 
the models studied and the details of the simulations and then introduce 
correlation functions that we have calculated to characterize the relevant 
relaxation processes in glass forming liquids. Finally 
we show our results and discuss the implications of these results in the 
context of bio-preservation. 

\begin{figure}[!h]
\hskip -1.50cm
\includegraphics[scale=0.34]{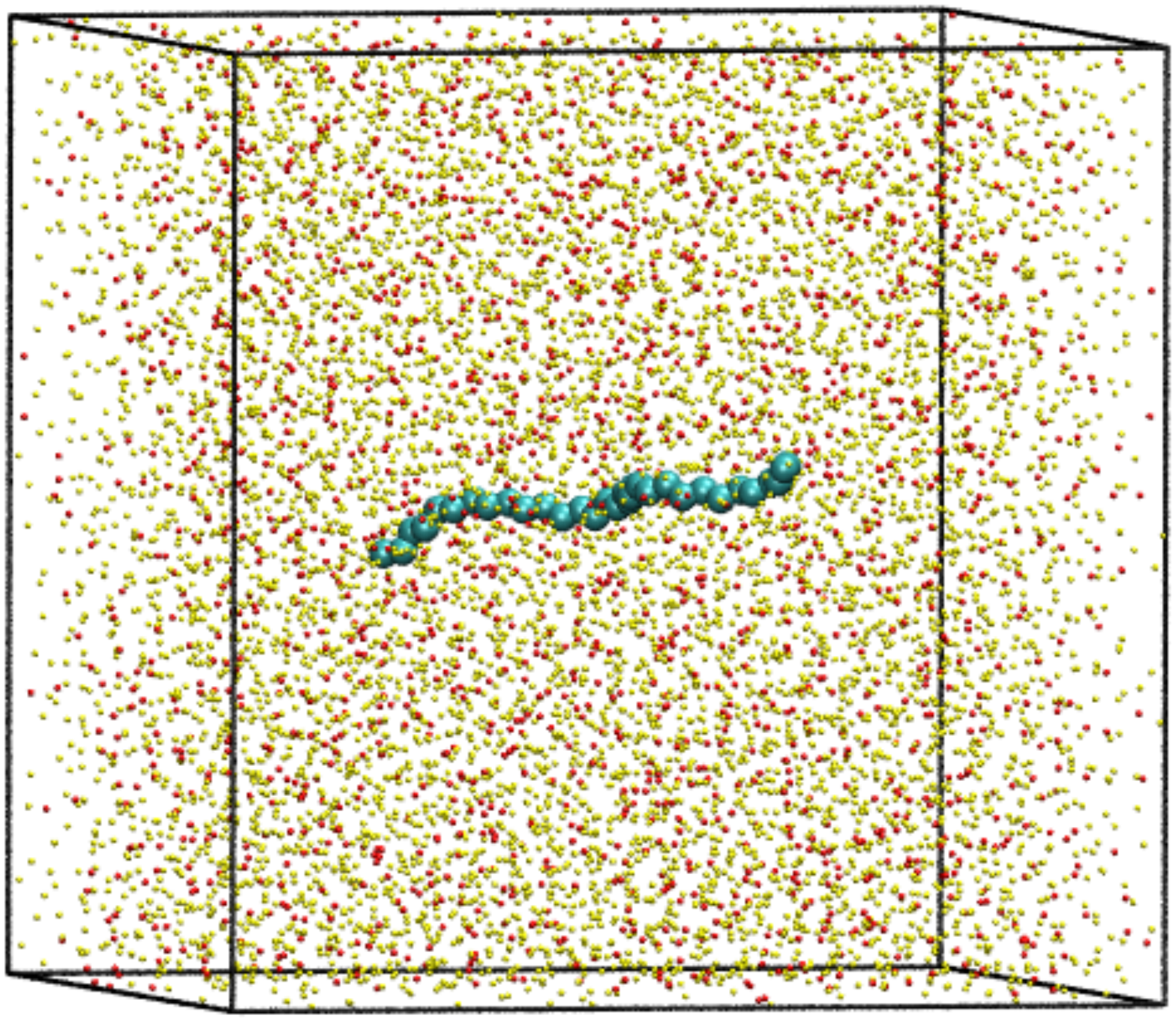}
\caption{Extended Configuration of the polymer inside the binary mixture. The 
actual size of the solvent glassy molecules have been scaled down in the 
figure for clarity.}
\label{in_confFig}
\end{figure}

\section{Models and Methods}
We have studied the well known Kob-Anderson 80:20 binary glass former Lennard-Jones 
mixture \cite{KA} as the solvent. The interaction potential in this model is 
given by,
\begin{equation}
V_{AB}(r) = 4\epsilon_{AB}\left[\left(\frac{\sigma_{AB}}{r}\right)^{12} - \left(\frac{\sigma_{AB}}{r}\right)^{6}\right]
\end{equation}
where $\epsilon_{AA} = 0.997$, $\epsilon_{AB} = 1.4955$, $\epsilon_{BB} = 0.4985$, 
$\sigma_{AA} = 0.34$, $\sigma_{AB} = 0.272$, $\sigma_{BB} = 0.2992$. The units of 
$\epsilon$ is kJ/mole and unit of  $\sigma$ is in nm (all are transformed to real 
unit in terms of Argon).

This binary mixture works as a solvent for a $32$ bead model polymer. The 
polymer model closely resembles that of Berne and coworkers \cite{BZZJACS2009}. 
The constituent polymer beads are connected to the covalently bonded neighbor 
by a harmonic potential, with an equilibrium bond length of  $0.153 nm$ (the 
same as CH2-CH2 bond length). The  angle between adjacent covalent bonds is 
represented by a harmonic potential, with an equilibrium angle of $111^o$ 
(the same as CH2-CH2-CH2 bond angle). The polymer 
is uncharged and the beads interact among themselves and with their environment 
via Lennard-Jones potentials.  The bead diameter is fixed at $\sigma_b=0.4nm$ 
and the bead-bead interaction is fixed at $\epsilon_b=11$ kJ/mol. Non-bonded 
interactions between a bead and its first and second nearest neighbors were 
excluded, and no dihedral interaction terms were included. The 
hydrophobic character of the chain can be tuned by varying the intermediate 
interaction between polymer beads and particles constituting glassy matrices 
using geometric combination rules. Specifically, the polymer-glass interaction 
potential is given by $\sqrt{\epsilon_p * \epsilon_{AA}}$ and $\sqrt{\epsilon_p * 
\epsilon_{BB}}$, where we have independently varied the value of $\epsilon_p=0.1, 
1.0$ and $3.0$ kJ/mol in separate simulations for tuning polymer-liquid 
interactions.  In essence, $\epsilon_p$ denotes the polymer contribution towards 
the inter polymer-glass interaction. On the other hand, the inter-polymer-glass 
interaction-range has been calculated by $\sqrt{\sigma_b * \sigma_{AA}}$ and 
$\sqrt{\sigma_b * \sigma_{BB}}$ . A cutoff of $1.2 nm$ was used to treat the 
nonbonding interactions and periodic boundary condition condition was implemented in 
all dimensions.  The
temperature range studied for this model is $50-120K$. Number of particles we 
have chosen for the binary mixture is $9600$ in a cubic box of dimension $6.8$ nm. 
The same average density was maintained throughout the simulations.

All the molecular dynamics simulations have been performed using $GROMACS \quad 5.1.4$ software. 
We have solvated the energy-minimized extended configuration of the polymer 
chain into the energy-minimized binary mixture for each case. The systems 
were first energy minimized by steepest descent algorithm and then equilibrated 
for $100$ ps at 260K  in NVT ensemble and then in NPT ensemble for $200$ ps. 
The systems were then annealed to desired temperatures at a cooling rate of 
$0.5$ K/ps and then subjected to a NPT equilibration for $20-1500$ ns depending
on the temperatures. Note that equilibration runs for each temperatures are at 
least $100\tau_{\alpha}$ or more longer. 
$\tau_{\alpha}$ is the $\alpha$ relaxation time 
(defined later) of the solvent glass forming liquid. Finally the systems 
were subjected to production run in NPT ensemble. The reference pressure 
for NPT simulations in last part of equilibration and production run was 
the average pressure obtained from the equilibrated binary mixture without 
the polymer. The integration time step used is $dt = 0.002$ ps. 
Berendsen and V-rescale thermostat has been used respectively during 
equilibration and production runs to maintain the average temperature. 
On the other hand Berendsen and Parrinello-Rahman barostat to keep 
pressure fixed during equilibration and production runs respectively.

\section{Results and Discussion}
\begin{figure}[!h]
\hskip -0.720cm
\centering
\includegraphics[scale=0.255]{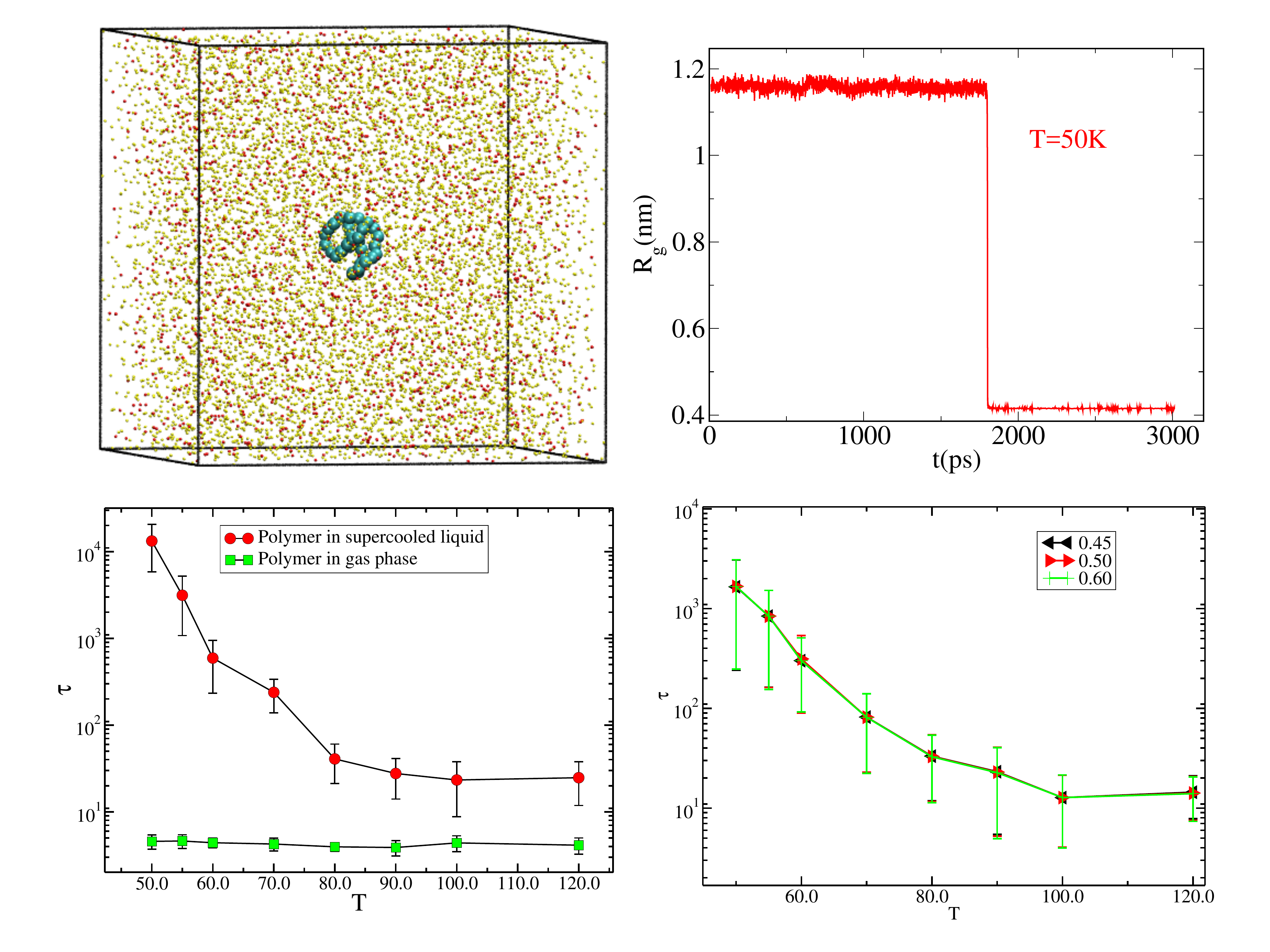}
\caption{Top left panel: Typical snapshot of the collapsed state of the polymer 
in the supercooled liquid. Top right panel: The time profile of the radius of 
gyration of the polymer chain at $T = 50K$. The step like change in the radius 
of gyration suggest that the collapsing transition for this polymer is very 
sharp. Bottom left panel: Timescale for degradation of polymer inside the 
supercooled Liquid and in Gas Phase. Notice the stark difference in the 
collapsing timescale of the polymer in gas phase and in the supercooled liquid
for different temperatures.  Bottom right panel: 
Temperature dependence of the collapsed time for three different cutoff radii 
of gyration for collapsed state of the polymer respectively $0.45$, $0.50$ 
and $0.60 nm$. }
\label{fig2label}
\end{figure}

All our simulations start with an extended  configuration of the polymer 
(Radius of gyration $R_G = 1.2 nm$) as shown in Fig.\ref{in_confFig} in a 
well equilibrated supercooled liquid state of the solvent mixture at different 
studied temperatures in the range $T\in [50K,120K]$ and we explore the 
transition of the polymer from extended to collapsed conformation during the 
course of the simulation. Time profile of Radius of gyration of the polymer 
is calculated (as shown in top right panel of Fig.\ref{fig2label} for 
$T = 50K$ ) to quantify the collapse-dynamics of the polymer and the 
collapse time, $\tau_c$ (defined later) is estimated by identifying the 
time of sharp transition from extended to collapsed conformation at 
different supercooling temperature.
This collapsing timescale is 
then compared with the intrinsic 
relaxation time scale ($\tau_{\alpha})$ of the glassy liquid. In 
left top panel of Fig.\ref{fig2label}, we have shown one such instance 
of the collapsed configuration of the polymer. The solvent binary 
supercooled liquid molecules are also shown by reducing their actual 
size for clarity.  

The choice of a large intra-bead interaction parameter ($\epsilon_b$ = 
$11 kJ/mol$) renders a strong propensity for the polymer-collapse and hence 
allows us to observe the collapse behavior of the polymer for entire range 
of temperature of interest $T\in [50K,120K]$ within simulation time scale.
So in gas phase the polymer collapses very quickly with a very low temperature 
dependence as shown in bottom left panel of Fig.\ref{fig2label} (green square 
symbols). The reason for choosing such polymer parameters is to explore whether 
dynamics of supercooled liquid can slave the dynamics of the polymer even 
when the polymer interacts weakly with the liquid compare to its own 
interaction strength. We also have shown a comparison of the collapsing 
timescale in the same panel (red circle) when the polymer is immersed in 
supercooled liquids with particular interaction (discussed in details later). 
The changes in the collapsing timescale compare to the gas phase timescale 
is really dramatic. This clearly proves why glassy matrices are chosen for 
bio-preservation.     

We have used three different $\epsilon_p$ value to control the interactions 
between glass molecules and polymer beads. The $\epsilon_p$ values for 
polymer-liquid interactions are $0.1, 1.0$ and $3.0 kJ/mol$. Before going 
in to discussing our main observations, we will briefly discuss how 
characterization of the supercooled liquid is done. Relaxation time 
is measured from the decay of a modified version of the two point 
density-density correlation function $Q(t)$, also known as overlap correlation
function \cite{KDS}. It is defined as
\begin{equation}
Q(t) = \sum_{i=1}^{N} w\left(|\vec{r}_i(0)-\vec{r}_i(t)|\right)
\end{equation}
where $\vec{r}_i(t)$ is the position of particle $i$ at time $t$, $N$ is the 
total number of particles. The window function $w(x) = 1$ if $x \le a$ and 
$0$ otherwise, where $a$ is a cut-off distance at which the the root mean 
square displacement (MSD) of the particles as a function of time exhibits a 
plateau before increasing linearly with time at long time. The precise choice 
of $a$ is qualitatively unimportant. This window function is chosen to remove
any de-correlation that might happen due to vibrational motions of the solvent 
particles inside the cages formed by their neighbours. In this study we have 
taken $a^2 = 0.006nm$. The relaxation time $\tau_{\alpha}$ is defined as
$\langle Q(t = \tau_{\alpha})\rangle = 1/e$, where $\langle\ldots\rangle$ 
refers to ensemble average.

The collapse time ($\tau_c$) of the polymer chain is obtained from the time 
dependence of the radius of gyration ($R_G$) of the polymer chain. In top 
panels of Fig.\ref{fig2label}, we have shown the $R_G$ as a function of time 
closed to the collapsing transition for  $T = 50K$ as an illustrative collapse 
profile. In all our analysis, we have considered the chain to be in the 
collapsed state when $R_G$ became $0.50nm$. 
In bottom right panel of Fig.\ref{fig2label}, we have shown the temperature 
dependence of the collapsed time for three different cutoff radius of gyration 
for collapsed state of the polymer as $0.45$, $0.50$ and $0.60 nm$. As evident, a different choice of the cut off radius gyration to 
define the collapsed state does not change the results qualitatively.

\begin{figure}[!h]
\centering
\includegraphics[scale=0.34]{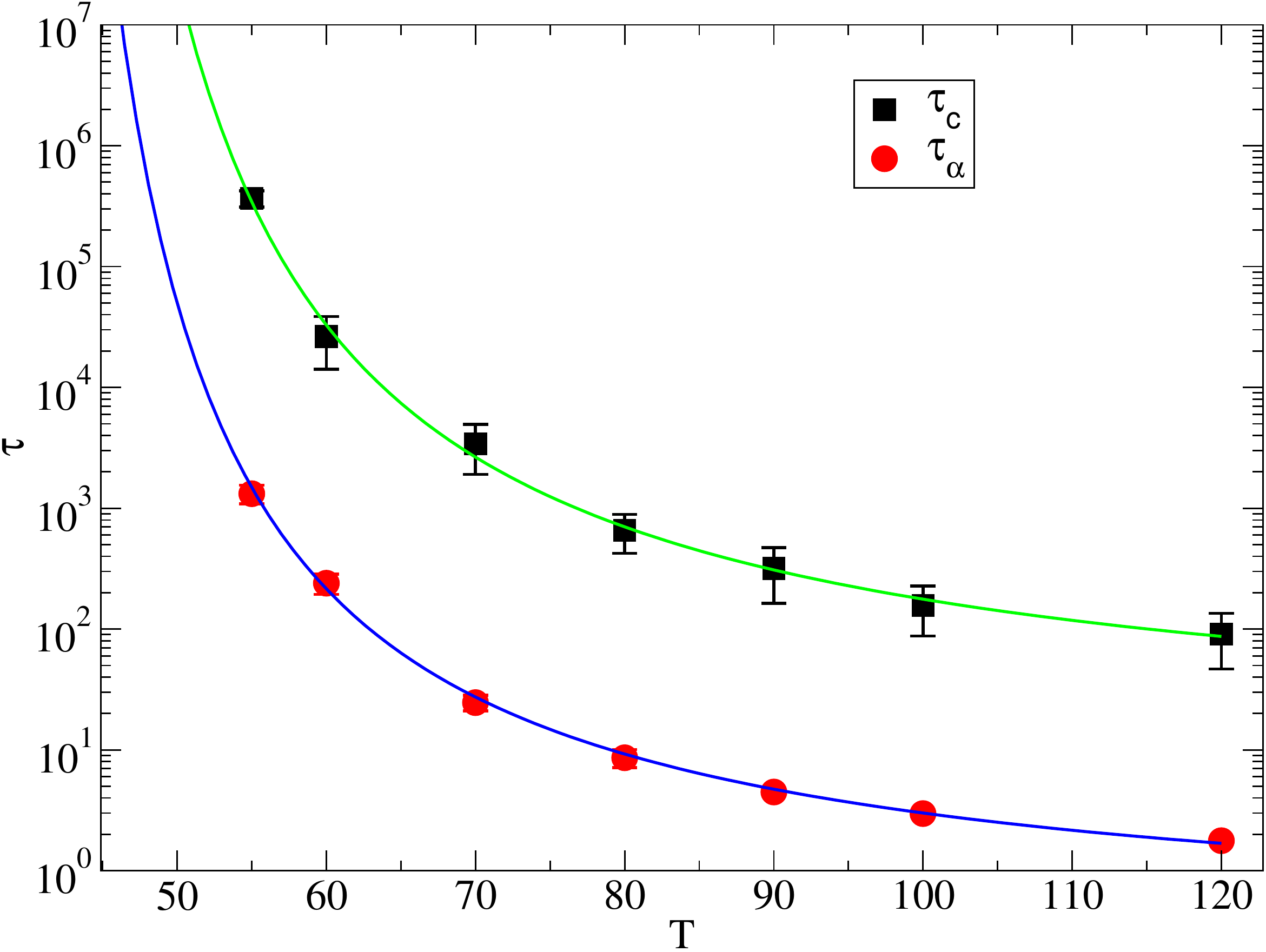}
\caption{Comparison of timescale for a polymer with $\epsilon_b$=11 kJ/mol and 
$\epsilon_p$=3 kJ/mol in the studied temperature of range. The lines are 
fit to the data with VFT formula (see text for details). The VFT divergence 
temperatures for both the timescales are found to be equal to $38K$, suggesting
a strong coupling between $\alpha$-relaxation time and collapsing time of the
polymer.}
\label{tauVsCollapsed3_11}
\end{figure}
Next we compare $\alpha$-relaxation time, $\tau_{\alpha}$ of the supercooled 
liquid and the collapse time, $\tau_c$ for a situation where the solvent 
supercooled molecules interacts somewhat strongly with the polymer chain 
molecules. Specifically the polymer-solvent intermediate interaction 
is tuned by using $\epsilon_p$ =3.0kJ/mol, keeping polymer bead-bead 
interaction fixed at $\epsilon_b$=11 kJ/mol. In Fig.\ref{tauVsCollapsed3_11} 
we have plotted collapse time of the polymer  along with the $\alpha$ 
relaxation time of the liquid for different temperatures for this particular 
choice of the parameter. It is clear that, at least in the studied temperature 
regime, $\tau_\alpha$ controls the degradation rate, supporting the 
``Vitrification Hypothesis''. One may infer that better stability needs 
larger value of $\tau_\alpha$ of the preservative. We have fitted both 
$\tau_{\alpha}$ and $\tau_c$ by Vogel-Fulcher-Tamman (VFT) formula \cite{vft},
defined as $\tau = \tau_0 \exp{\left(A/(T-T_0)\right)}$, where $\tau_0$, $A$ 
and $T_0$ are free parameters. $T_0$ is known as VFT divergence temperature 
and is very closed to the Kauzmann Temperature \cite{kauz}. The divergence 
temperatures for both $\tau_{\alpha}$ and $\tau_c$ are found to be close 
to $38K$ suggesting a strong coupling between the dynamics of supercooled 
liquid and the collapsing dynamics of the polymer chain. Note that the 
polymer chain collapse very rapidly in gas phase, whereas its dynamics now 
is slaved to the dynamics of the solvent glassy liquids.

\begin{figure}[!h]
\centering
\includegraphics[scale=0.33]{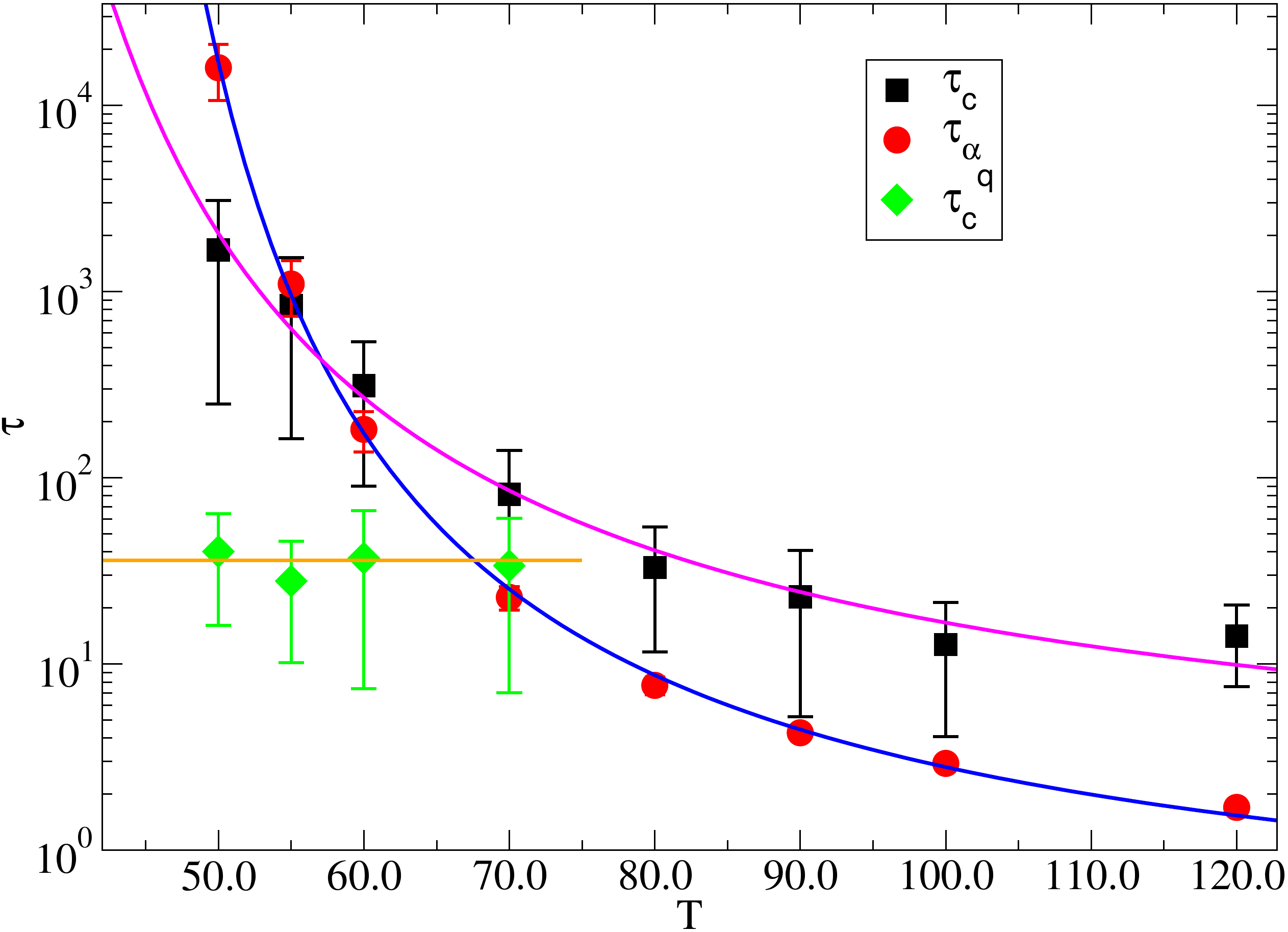}
\caption{Comparison of timescale for a polymer with $\epsilon_b$=11 kJ/mol 
and $\epsilon_p$=3 kJ/mol for the temperature of range of interest. The lines
are the fit to the VFT formula (see text). $\tau_c^{q}$ is the collapsing 
time obtained in the quench studies (see the text for further details).}
\label{tauVsCollapsed01_11}
\end{figure}

Next we look at the other extreme in which we choose an interaction 
parameter such that the polymer chain interacts very weakly with the 
solvent liquid molecules. We choose the value of $\epsilon_p$ 
contributing to solvent-polymer interaction to 
be $0.1kJ/mol$, so in this limit polymer dynamics will be mainly 
affected (if at all) by the crowding effect of the solvent glassy 
molecules. In Fig.\ref{tauVsCollapsed01_11}, we show the temperature 
dependence of the two timescales and surprisingly, they cross each 
other at some intermediate temperature, $T \sim 55K$ in this case. 
The corresponding VFT fits also suggest that the extrapolated divergence 
temperatures are very different from each other. This is now in contrast 
with our previous observation, where both the timescales are more or 
less proportional to each other. This new result suggests that only 
$\alpha$-relaxation time is not the main controlling parameter, 
especially at lower temperatures, where a much faster relaxation process, 
probably $\beta$-relaxation process seems to play a role in the dynamics 
of the polymer chain. This is in complete agreement with recent 
experimental observations \cite{CiceroneDouglasSoftMatter2012, CiceroneDouglasBioPhyJ2004}, where it is suggested that protein 
preservation in sugar glasses is directly linked to high frequency
$\beta$-relaxation process as protein stability seems to increase 
almost linearly with $\tau_{\beta}$ when $\tau_{\beta}$ is increased by 
adding anti-plasticizing additives. These additives are found to increase 
the $\beta$-relaxation time even though it decreases $\alpha$-relaxation 
time \cite{review1}.

In a bid to further understand whether it is the glassy dynamics that is 
slowing down the collapsing dynamics of the polymer chain, we performed 
quenching studies in which we decrease the temperature of the supercooled
liquid rapidly from its initial equilibrium temperature. It is well known 
that if we quench a glass forming liquid to low temperature then it shows 
aging and initially it relaxes almost at the same timescale as that of 
the initial temperature from which it is quenched. The relaxation time 
then gradually increases with increasing waiting time. Now, if the dynamics 
of the polymer chain is slaved to the dynamics of the supercooled liquid, 
then if we quench the whole system, the polymer should still be able to 
collapse in a timescale which is almost same as the initial temperature 
from which it is cooled. 
In Fig.\ref{tauVsCollapsed01_11}, we show that collapse time (referred 
here as $\tau_c^q$) seems 
to depend on the initial temperature ($T = 120K$) from which it is 
quenched, irrespective of the final temperatures (green diamonds, 
$T = 70, 60, 55, 50K$ respectively). In all these quench studies, 
the equilibrium collapse time is many orders of magnitude larger 
than the time obtained if the system is quenched to these 
temperatures from high temperature. This observation seems to 
corroborate with an old experimental finding \cite{mazurSchmidt}, 
where it was noted that survival probability of frozen and thawed 
yeast is orders of magnitude more if it is cooled very slowly.

\begin{figure}[!h]
\centering
\includegraphics[scale=0.41]{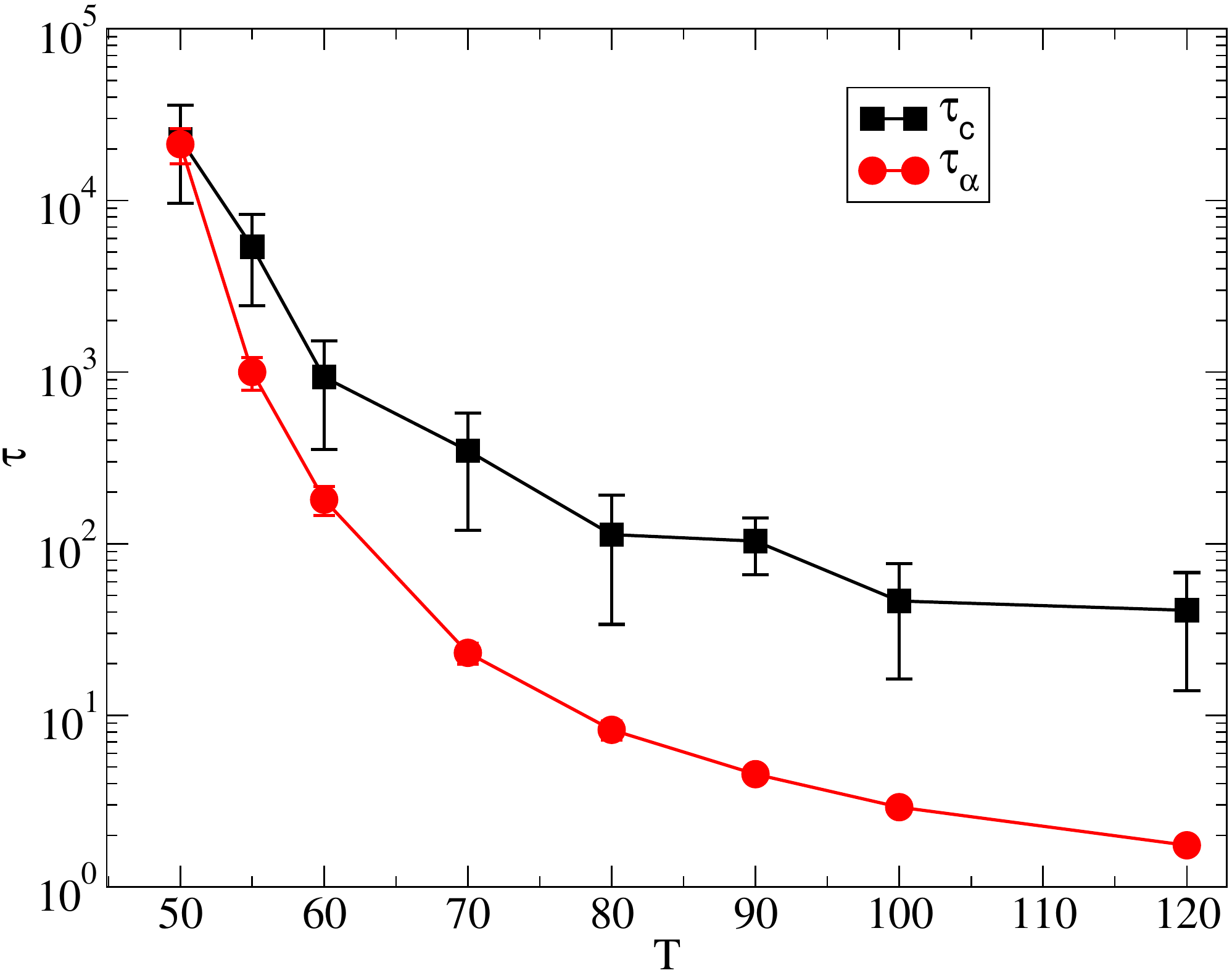}
\caption{Temperature dependence of $\alpha$-relaxation time and
collapse time for the case of $\epsilon_p$ equal 
to $1.0$ kJ/mol. The cross over temperature is now shifted to lower
temperature compare to the case when $\epsilon_p$ was $0.1$ kJ/mol.}
\label{tauVsCollapsed1_11}
\end{figure}

We then increase the polymer-solvent interactions a bit more by 
increasing the value to an intermediate value $\epsilon_p$ to 
$1kJ/mol$ to see whether these two timescales still cross each other 
at an accessible temperature range. In Fig.\ref{tauVsCollapsed1_11}, 
we indeed see the crossing of these two timescales but now at a 
temperature lower than that observed in the previous case when the 
polymer-solvent interaction is $0.1kJ/mol$. With this particular 
parameter, the crossover temperature moves to $50K$. Thus we can 
expect that at some intermediate parameter of $\epsilon_p$ =$1-3$ 
kj/mol guiding the polymer-solvent interaction, the 
collapsing dynamics of the polymer chain will be completely 
controlled by $\alpha$-relaxation and below that short time $\beta$
relaxation will also be important. At this point we can not rule 
out the other possibility of cross over of these two timescales at 
all parameter range, as our conclusions are based on the 
extrapolation done using VFT formula. 

In conclusion, we have shown that dynamics of supercooled glass 
forming liquids play a major role in controlling the collapsing 
dynamics of a polymer chain at various temperature. At certain 
polymer-solvent interaction strength, the polymer can be completely 
slaved to the long time $\alpha$-relaxation of the glassy liquid,
on the other hand at low polymer-solvent interaction strength, 
at which the polymer is passive to the liquid and only packing 
of the solvent molecules around the polymer molecule is relevant, 
both short time $\beta$ and long time $\alpha$ relaxations play 
intricate role at different temperature regimes. We also have 
shown that coupling between the solvent dynamics and polymer 
becomes weak if one does quenches from high temperatures due 
to aging in the glassy liquids. This suggests that flash freezing 
might not be a good method if one wants to preserve a biomolecules 
in glassy matrix. Thus ``Vitrification Hypothesis'' although might 
be valid for some biomacromolecules, need serious revision to include 
the effect of shorter time scale processes like $\beta$-relaxation 
in order to better understand bio-preservation in glassy sugar 
matrix. In a recent work \cite{KwonPRL2017}, it is shown that 
reaction kinetics of polymer collapsing dynamics depends on 
viscosity of supercooled liquids with a fractional power. This 
again supports our findings reported in this work very strongly. 
Finally, in our model studies all complicated interactions 
like hydrogen bonding and complex structural aspects of the 
biomolecules are not incorporated, thus it will be important to 
do further studies to understand how these different parameters 
influence the results reported here.   

\acknowledgements{
We would like to thank Walter Kob for suggesting us to do the aging 
study and Sashi Thutupalli for bringing Ref.\cite{mazurSchmidt} to 
our notice. We would also like to thank Srikanth Sastry, Juergen 
Horbach and Surajit Sengupta for many useful discussions.}

\end{document}